\documentclass[12pt,a4paper,reqno]{article}

\usepackage{amsmath}

\usepackage{graphicx}

\newcommand{\pa}{\hspace{0.5em}}
\newcommand{\speech}{\medskip\setlength{\hangindent}{2cm}\noindent}

\newcommand{\X}{\sigma_x}
\newcommand{\Y}{\sigma_y}
\newcommand{\Z}{\sigma_z}
\renewcommand{\>}{\rangle}

\begin{document}

\title{Alice and Bob get away with it: A playlet}
\author{Anthony Sudbery\\[10pt] \small Department of Mathematics,
    University of York,\\[-2pt] \small Heslington, York, England YO10 5DD\\
    \small  Email:  as2@york.ac.uk}
\date{}
\maketitle

\begin{abstract}
Alice and Bob use Aravind's version of the Bell-Kochen-Specker theorem to fend off awkward questions about what exactly they were doing in Amsterdam last week.
\end{abstract}

\section{Introduction}

Jacobs and Wiseman\cite{crime} use a story about a robbery to illustrate Bell's theorem, discussed in Mermin's version\cite{Mermin-GHZ} based on the Greenberger-Horne-Zeilinger (GHZ) state of three qubits \cite{GHZ}. They ask if a story can be constructed around Aravind's recent version \cite{Aravind} of the Bell-Kochen-Specker theorem \cite{Bell:KS, KochenSpecker}. Aravind's proof is a particularly clear and simple demonstration of the incompatibility between quantum mechanics and our intuitive classical ways of thinking about the physical world. He shows how an entangled state of four qubits (for example, the spin states of four electrons) can be used by two separated people, with no means of communication, to perform a joint task that would appear to be impossible without communication. This and similar applications of quantum entanglement have become known as ``pseudo-telepathy."\cite{telepathy} Although it cannot be used as an instantaneous method of communication, pseudo-telepathy is possibly even closer to the science-fiction idea of telepathy than quantum teleportation\cite{teleport} is to its science-fiction namesake. It is therefore an attractive subject for describing the peculiarities of quantum information to students and the general public, and one that is particularly well suited to illustration in story form. 

Such a story was told in a little play which was part of the Merchant Adventurers' Science Discovery lecture, a public lecture given in York in March 2006. The script of this playlet is given in Sec.~II. The theory of the device used by Alice and Bob is explained in Sec.~III.

\section{The script: Alice and Bob Get Away With It}

The play was first performed in the Merchant Adventurers' Hall, York, on 7 March 2006. The part of Alice was played by Chris Higgins, that of Bob by Ged Murray. The Lecturer was Tony Sudbery.

The lecturer has just explained the EPR experiment and why it led Einstein, Podolsky and Rosen to conclude that each qubit has definite properties which determine the results of all possible experiments on the qubit.

\speech Lecturer: \pa But Einstein was wrong. John Bell showed that there are situations where we can't understand the effects of entanglement by saying that each side has its own independent properties. And this is what gives entanglement its power: it gives rise to teleportation, \cite{teleport} telepathy, \cite{telepathy} and ---

\speech BOB (\emph{coming up from audience}): \pa Hey, wait a minute. You don't expect us to believe this crap, do you? Parallel universes, telepathy, teleportation\ldots We came here for a serious science lecture, and you give us science fiction. This is the Merchant Adventurers' Hall, you know, not the City Screen.

\speech Lecturer: \pa It's Bob!

\speech Alice (\emph{coming up from audience}):
Just you be quiet, Bob Murray, and hear what he's got to say. You might learn something if you give your ears a chance.

\speech Lecturer: \pa Alice! It's all coming true.

\speech Alice: \pa Did you mention telepathy? That's the only thing that would help us in our current problem.

\speech Lecturer: \pa What problem is that?

\speech Alice: \pa We've just got to get our stories straight about last week.

\speech Bob: \pa Won't you ever give up? I've proved to you that we're never going to be able to do that the way you want to.

\speech Alice (\emph{sarcastically}): \pa Mathematically, I suppose.

\speech Bob: \pa Well, yes.

\speech Lecturer: \pa Sounds interesting. Tell me more.

\speech Alice: \pa We're accountants, and last week we were in Amsterdam looking at the accounts of three branches of our client's company. Their head office thinks there's some funny business going on. Tomorrow we'll be debriefed separately. A senior accountant will look at one of the branches with Bob, and our head of IT will look at the email traffic between the branches on one of the three days that we were there. We've got all the data, but we'll have to tell them exactly which branch we visited on which day.

\speech Lecturer: \pa What's the problem? Why don't you just tell them what you did?

\medskip

\emph{Bob and Alice exchange looks.}

\speech Lecturer: \pa I see. I didn't know accountants had such interesting lives. Well, why can't you get together beforehand and decide on your story?

\speech Alice: \pa The thing is, it will make life much easier if we can say we did things in a particular way. On each day we should visit at least one branch; maybe just one, but if we go on to a second one we must also visit the third on the same day, otherwise the first two will have time to discuss things and warn the third one. We don't have to have inspected all three branches, but if we do visit one we must go twice -- after the first visit we would look up our records in the evening and then go back to check the next day.

\speech Bob: \pa So when they ask Alice what we did on one of the three days, we want to say we went to either one or three branches; and when they ask me what we did about one of the branches, I want to say that either we didn't visit it or we visited it on two days. But we can't put together a schedule that does that! If we make either one or three visits on each of three days, then altogether we would have made an odd number of visits (three odd numbers add up to an odd number), but if we visit each branch either no times or twice, that makes an even number of visits in total.

\speech Lecturer: \pa Surely your bosses can work that out just as well as you can. They can't expect you to have done the impossible.

\speech Alice: \pa No, but if we don't do it this way they will want to probe a lot further. It would save a lot of awkward questions if we can be sure that whatever they ask us, we have an answer that fits these rules. But we've got to be sure that when they compare notes our stories are compatible --- we must both say that we went to the same branch on the same day.

\speech Lecturer: \pa I'm not sure that I really understand why you want to do this --- I don't think I want to know. But yes, entanglement can do it for you. As if you weren't entangled enough already. 

\setlength{\hangindent}{2cm} \hspace{2cm} Let me see if I've got it straight. You have to convince your bosses that you undertook a schedule of visits to the three different branches, $A$, $B$ and $C$, on the three days that you were in Amsterdam. So you could fill in the grid in Table~\ref{grid} by putting a tick in the appropriate square if you visited a particular branch on a particular day. Alice will be asked to fill in one of the rows, and she wants to do so with an odd number of ticks; Bob will have to fill in one of the columns, and he wants to put an even number of ticks. But you want to be sure that you both put the same thing in the square where Alice's row and Bob's column intersect. Right?

\begin{table}[h]
\begin{tabular}{l|p{2cm}|p{2cm}|p{2cm}|}
& Branch A & Branch B & Branch C \\
\hline
Tuesday &&& \\
\hline
Wednesday &&& \\
\hline
Thursday &&& \\
\hline
\end{tabular}
\caption{\label{grid} The grid to be filled in by Alice and Bob}
\end{table}

\speech Alice and Bob (\emph{together}): Right.

\speech Lecturer: You're right, you can't do that with a complete schedule. What you need is the Quantum Entangulator. Here you are -- you each have a handset, cunningly disguised as a soft toy (a cat, as it happens). Each of these contains a pair of electrons, and a measuring device which will do an experiment on the electrons and translate the result into an answer to the question you have been asked, obeying the restrictions you've described. The electrons are all entangled together in such a way that the answers are guaranteed to fit together.

\setlength{\hangindent}{2cm} \hspace{2cm} Let's test it. Switch them on. Now, Alice, suppose you're asked which branches you visited on Tuesday. Press button 1 to find out what you should answer.

\speech Alice: \pa It says B.

\speech Lecturer: \pa And Bob, suppose you're asked on which days you visited branch B. Press button B to find your answer.

\speech Bob: \pa Tuesday and Wednesday.

\speech Lecturer: \pa You see! They match up. And they always will. As Bob proved, the answers cannot be programmed in advance; there is no physical connection between the two ends of the Entangulator --- no wires, no phone line, no radio; but they are guaranteed to give compatible answers.

\speech Bob: \pa Fantastic! The university should patent this!

\speech Lecturer: \pa Not our idea, I'm afraid, Bob. \cite{Aravind}

\speech Alice: \pa But have they realized what they've got here? When I got my diary entry from my end of the Entangulator, Bob's end immediately got the information about the branch they were asking about. No radio or anything --- so it was going faster than light! You've really beaten Einstein in a big way here. It may not matter now, but think what it will mean in the future! Like, when spaceships are trying to send messages over huge distances --- beating the speed of light will be really important then.

\speech Lecturer: \pa Calm down, Alice. I agree it looks good --- but how are you going to use it?

\speech Alice: \pa Well, I put my message in the entangulator thingy, and it immediately comes out at Bob's end.

\speech Lecturer: \pa But Alice, when you use the entangulator tomorrow morning, you'll never put anything into it. You only get something out of it --- and you have no control over what that is. Bob gets a corresponding thing out, but there's no way you can make that be the message you want to send.

\setlength{\hangindent}{2cm}\hspace{2cm} Other people have suggested that something was moving faster than light in the entangulator, but they really ought to know better.\cite{speedqinfo} When you activate your side of it, it seems to change Bob's side, it's true, because it changes the possible answers that can be flashed up on his screen; some answers become impossible. But Bob doesn't know that. Just because he doesn't get a particular answer on a single occasion, he can't deduce that it's become impossible. He doesn't get any information until you've told him independently what answer you got.

\setlength{\hangindent}{2cm}\hspace{2cm} In general, there's a theorem that entanglement can never be used by itself to send signals. \cite{nosignalling} In particular, it can't communicate faster than light. So if you want to send Bob a kiss in the quantum wonderland, you'll just have to do it the old-fashioned way.

\section{The Explanation}

The entangulator contains four qubits, two in Alice's cuddly toy and two in Bob's, in the state
\begin{align}
|\Psi\> & = |00\>_A|00\>_B + |01\>_A|01\>_B + |10\>_A|10\>_B +
|11\>_A|11\>_B,\\
\noalign{\noindent which can be written as}
\label{4qubit}
|\Psi\> & = |E_1\>_A|E_1\>_B + |E_2\>_A|E_2\>_B +
|E_3\>_A|E_3\>_B + |E_4\>_A|E_4\>_B
\end{align}
where $|E_1\>,\ldots |E_4\>$ is any orthonormal basis of two-qubit states that have real coefficients when expanded in the basis $|00\>,|01\>,|10\>,|11\>$. Hence if Alice measures her two qubits in any such basis and obtains the result $|E_i\>$, Bob's qubit will be projected into the same state $|E_i\>$. The measurements made by Alice and Bob are shown in Table~\ref{qubit}:

\begin{table}[h]
\begin{tabular}{l|p{2cm}|p{2cm}|p{2cm}|}
& Branch A & Branch B & Branch C \\
\hline
Tuesday & $I\otimes \Z$ & $\Z\otimes I$ & $-\Z\otimes \Z$ \\
\hline
Wednesday & $\X\otimes I$ & $I\otimes \X $& $-\X\otimes \X$ \\
\hline
Thursday & $\X\otimes \Z$ & $\Z\otimes \X$ & $-\Y\otimes \Y$ \\
\hline
\end{tabular}
\caption{\label{qubit} The measurements made by Alice and Bob}
\end{table}

\noindent Each row and each column of Table~\ref{qubit} contains a set of three commuting two-qubit observables. Each row therefore determines a basis of two-qubit states, namely the simultaneous eigenvectors of the observables in that row. The transformation matrices between these bases are purely real, so any of the bases could serve as $|E_1\>,\ldots,|E_4\>$ in Eq.~\eqref{4qubit}.

When she is asked about a particular day, Alice measures the commuting observables in the corresponding row of the table, and converts the result into an answer to the question about which branches they visited that day: if the observable under branch $X$ is measured to have the value $-1$, she says that they visited branch $X$ that day; if the value is $+1$ she says they didn't visit it. This measurement projects Bob's pair of qubits onto the same simultaneous eigenstate. When he is asked when they visited branch $X$, he measures the commuting observables in the corresponding column, and converts the result into an answer to the question of when they visited this branch, in a similar way to Alice. If the observable in the Tuesday row has the value $-1$, he says they visited branch $X$ on Tuesday; otherwise they didn't. Because his state is now an eigenstate of the observable that is common to his set and Alice's, he is bound to obtain the same value for this observable. He therefore gives the same answer as Alice to the question in the square common to his column and her row. Thus, they both agree as to whether they visited his branch on her day.

The product of the observables in each row is $-I\otimes I$, while the product of the observables in each column is $I\otimes I$. The result of Alice's measurement therefore includes $-1$ an odd number of times --- she says that they visited an odd number of branches on each day --- while the result of Bob's measurement includes $-1$ an even number of times -- he says that they visited each branch an even number of times. Whatever they were doing in Amsterdam, they get away with it.


\end{document}